\documentclass{article}
\usepackage[ruled,vlined]{algorithm2e}
\usepackage{array}
\usepackage{graphics}
\usepackage{graphicx}
\usepackage{makecell}
\usepackage{multicol}
\usepackage{multirow}
\usepackage{physics}
\usepackage{siunitx}
\usepackage{tabularx}

\usepackage[preprint, nonatbib]{neurips_2020}
\usepackage[utf8]{inputenc} 
\usepackage[T1]{fontenc}    
\usepackage{hyperref}       
\usepackage{url}            
\usepackage{booktabs}       
\usepackage{amsfonts}       
\usepackage{nicefrac}       
\usepackage{microtype}      

\newcommand{\absdiv}[1]{
  \par\addvspace{.5\baselineskip}
  \noindent\textbf{#1}\quad\ignorespaces
}

 \graphicspath{ {./figures/} }

\title{Data-Limited Tissue Segmentation using Inpainting-Based Self-Supervised Learning}

\author{
    Jeffrey Dominic\\
    Department of Radiology\\
    Stanford University\\
    \And
    Nandita Bhaskhar\\
    Department of Electrical Engineering \\
    Stanford University\\
    \AND
    Arjun D. Desai\\
    Department of Radiology\\
    Department of Electrical Engineering\\
    Stanford University\\
    \And
    Andrew Schmidt\\
    Department of Radiology\\
    Stanford University\\
    \AND
    Elka Rubin\\
    Department of Radiology\\
    Stanford University\\
    \And
    Beliz Gunel\\
    Department of Radiology\\
    Department of Electrical Engineering\\
    Stanford University\\
    \AND
    Garry E. Gold\\
    Department of Radiology\\
    Stanford University\\
    \And
    Brian A. Hargreaves\\
    Department of Radiology\\
    Department of Electrical Engineering\\
    Department of Bioengineering\\
    Stanford University\\
    \AND
    Leon Lenchik\\
    Department of Radiology\\
    Wake Forest School of Medicine\\
    \And
    Robert Boutin\\
    Department of Radiology\\
    Stanford University\\
    \AND
    Akshay S. Chaudhari\\
    Department of Radiology\\
    Department of Biomedical Data Science\\
    Stanford Cardiovascular Institute\\
    Stanford University
}

\begin{document}

\maketitle
\newpage

\begin{abstract}
    \absdiv{Purpose}
    To evaluate the efficacy of two self-supervised learning (SSL) methods (inpainting-based pretext tasks of context 
    prediction and context restoration) for 
    medical image segmentation in label-limited scenarios, and to investigate the effect of 
    implementation design choices for SSL on downstream segmentation performance.
    
    \absdiv{Methods}
    Manual segmentation labels were created for 3D knee MRI and 2D abdominal CT datasets. Multiple versions of self-supervised U-Net models were trained to segment tissues in both
    datasets, each using a different combination of design choices and pretext tasks 
    to determine the effect of different design choices on segmentation performance.
    The combination of these design choices that resulted in 
    the most significant improvement in Dice score 
    over supervised learning for both datasets was used to train an 
    optimally trained model for segmentation. 
    This model was pretrained on different amounts of unlabeled data to determine the 
    effect of pretraining dataset size on segmentation performance. 
    The highest performing models from this experiment were compared with 
    baseline supervised models for computing clinically-relevant metrics
    in label-limited scenarios.
    
    \absdiv{Results}
    SSL pretraining with 
    context restoration using 32x32 patches and Poission-disc sampling, transferring only the pretrained encoder weights, and fine-tuning immediately 
    with an initial learning rate of 1e-3 provided
    the most benefit over supervised learning for MRI and CT tissue 
    segmentation accuracy (p$<$0.001). For both datasets and most 
    label-limited scenarios, 
    pretraining using the maximum amount of unlabeled images resulted in better segmentation performance than pretraining using 
    only the training set (p$<$0.05). SSL models pretrained with this amount 
    of data also outperformed 
    baseline supervised models in the computation of clinically-relevant metrics 
    in scenarios with very low amounts of labeled data, especially for challenging classes to segment such as intramuscular adipose tissue 
    on CT images and patellar cartilage on MR images.
    
    \absdiv{Conclusion}
    We demonstrate how SSL can overcome paucity of labels for improving tissue segmentation by using unlabeled datasets. 
\end{abstract}

\section{Introduction}

Segmentation is an essential 
task in medical imaging that is common
across different imaging modalities and fields 
such as
cardiac, abdominal, and musculoskeletal imaging, amongst others
\cite{campello2021multi, kavur2021chaos, desai2021international}. Deep learning (DL) has enabled high 
performance on these challenges, but the power-law relationship between algorithmic performance and the amount of 
high-quality labeled training data fundamentally limits robustness and widespread use \cite{desai2019technical}.

Recent advances in self-supervised learning (SSL) provide an opportunity to reduce the annotation burden for 
deep learning models. In SSL, a model is first pretrained on a ``pretext" task, during which  
unlabeled images are perturbed and 
the model is trained to predict or correct the perturbations.
The model is then fine-tuned for downstream tasks.
Previous works have shown that such models can achieve high performance even when fine-tuned on only a 
small labeled training set \cite{pathak2016context, chen2019self, noroozi2016unsupervised}.
While most SSL models in computer vision have been used for the downstream task of image classification, segmentation comparatively remains an under-explored task \cite{jing2020self}.

\begin{figure}[ht]
    \centering 
    \includegraphics[width=\textwidth]{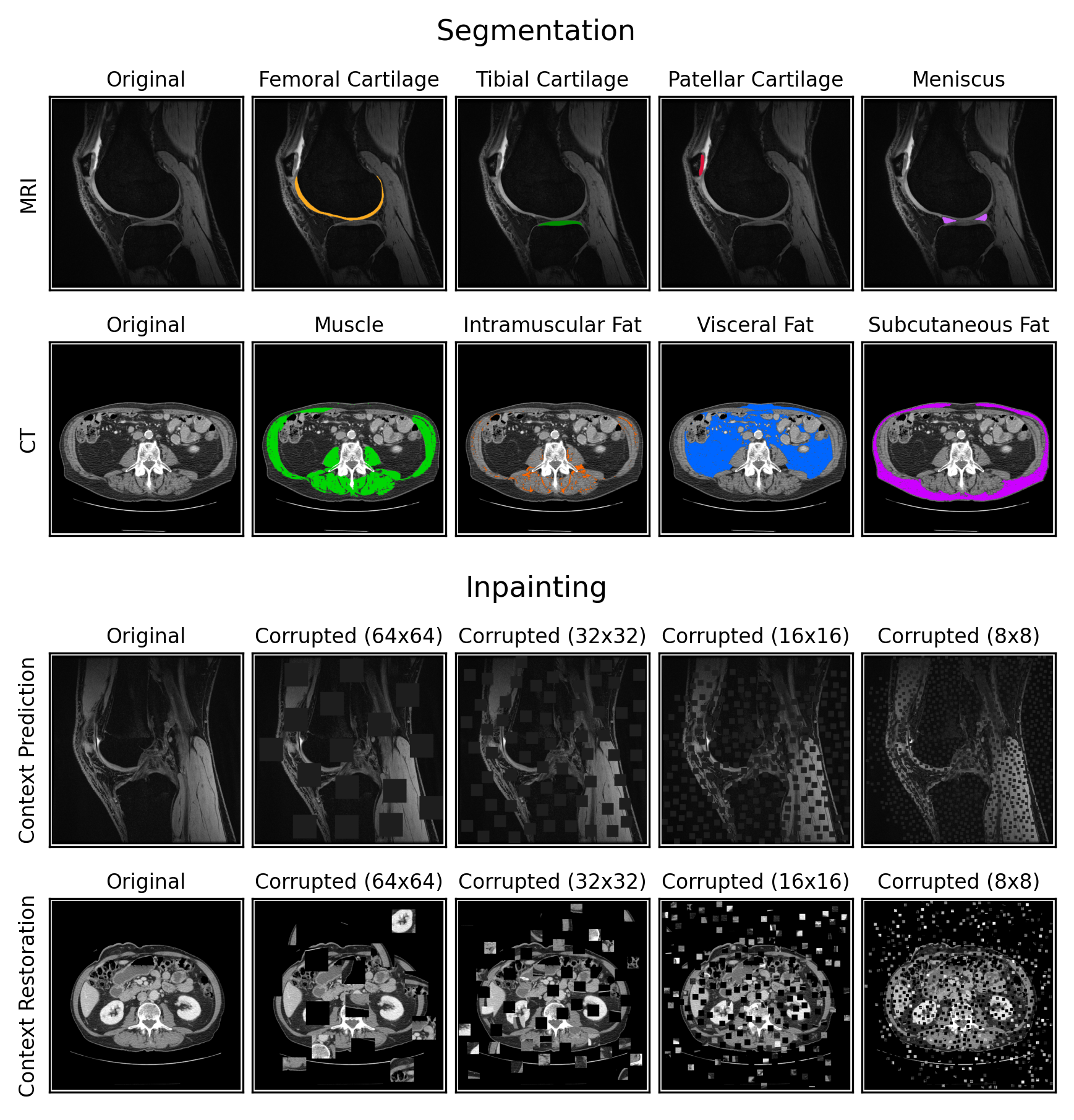}
    \caption{Example ground truth segmentations for the MRI and CT datasets (both with dimensions 512x512), and 
    example image corruptions for context prediction (zero-ing image patches) and context restoration 
    (swapping image patches). Since image corruption happens after
    normalization, the zero-ed out image patches for context prediction were actually replaced with the 
    mean of the image. The ``Inpainting" 
    section depicts image corruptions with four different patch sizes: 
    64×64,
    32×32, 16×16, and 8×8. The locations of these patches were determined using Poisson-disc sampling to 
    prevent randomly overlapping patches.}
    \label{fig:example_images}
\end{figure}

In this work, we systematically evaluate the efficacy of SSL for medical image segmentation across two domains -- 
MRI and CT.
We investigate ``context prediction" \cite{pathak2016context} and ``context restoration" \cite{chen2019self}, two 
well-known and easy-to-implement archetypes of restoration-based pretext tasks that produce image-level 
representations during pretraining
for eventual fine-tuning.
Context prediction sets pixel values in random image patches to zero, 
while context restoration randomly swaps pairs of image patches within an image while maintaining the distribution of pixel values 
(Figure \ref{fig:example_images}). For both tasks, the model needs to recover the original image given the corrupted image,
a process we refer to as ``inpainting".
We consider these two tasks because they maintain same input-output sizes, akin to segmentation.
We hypothesize that such pretext tasks allow construction of useful, image-level representations 
that are more suitable for downstream segmentation.

While context prediction and context restoration have been proposed before, the effects of the large space 
of design choices for these two pretext tasks, 
such as patch sizes for image corruption and learning rates for transfer learning, are unexplored. 
In addition, prior works exploring SSL for medical 
image segmentation have primarily focused on the accuracy of segmentation using metrics such as 
Dice scores \cite{chen2019self, chaitanya2020contrastive}, but have not investigated if SSL can improve 
clinically-relevant metrics, such as T2 relaxation times for musculoskeletal MRI scans and mean Hounsfield Unit (HU) values for CT scans. These metrics can provide biomarkers of biochemical changes in tissue structure prior to the onset of gross morphological changes \cite{T2mri2020, HUCT2022}. Furthermore, within the context of empirical data scaling laws in DL, past SSL works have rarely explored 
benefits of increasing the number of unlabeled images during pretraining \cite{goyal2019scaling}.  Characterizing the efficiency of SSL methods with unlabeled data can lead to more informed decisions regarding data collection, an important practical consideration for medical image segmentation. 
In this work, we address the above gaps by 

(1) investigating how different design choices in 
SSL implementation affect the quality of the pretrained model, 

(2)
calculating how varying 
unlabeled data extents 
affects SSL performance for downstream segmentation, 

(3) quantifying our results using clinically-relevant 
metrics to investigate if SSL can outperform supervised learning in label-limited scenarios,

(4) evaluating where SSL can improve performance, across different extents of labeled training data availability, and

(5) providing detailed analyses, recommendations, and open-sourcing our code to build optimal SSL models for medical image segmentation.

\section{Materials and Methods}
        
    \subsection{Datasets}
        \label{sec:datasets}
        
        \subsubsection{MRI Dataset}
        \label{sec:mri_dataset}
        
        We used 155 labeled knee 3D MRI volumes (around 160 slices per volume) from the SKM-TEA dataset 
        \cite{desai2021skm} and 86 unlabeled volumes (
        around 160 to 180 
        slices per volume), each with slice dimensions of 512x512 (other scan parameters 
        in \cite{desai2021skm}). 
        All volumes were acquired using a 5-minute 
        3D quantitative double-echo in steady-state 
        (qDESS) sequence,  which has been used for determining morphological and 
        quantitative osteoarthritis
        biomarkers and for routine diagnostic knee MRI \cite{chaudhari2018five, Chaudhari2018jmri, eijgenraam2020time, chaudhari2021diagnostic}. The 
        labeled volumes included manual segmentations for the femoral, tibial, and patellar cartilages, and the meniscus. 
        The labeled volumes were split into 86 volumes for 
        training, 33
        for validation, and 36 for testing, following the splits prescribed in \cite{desai2021skm}. 
        The 86 training volumes were further split into additional
        subsets, consisting of
        50\% (43 volumes), 25\% (22 volumes), 10\% (9 volumes), and 5\% (5 volumes) training data,
        to represent label-limited scenarios. 
        All scans in smaller subsets were included in larger subsets.
        
        \subsubsection{CT Dataset}
        \label{sec:ct_dataset}
        The 2D CT dataset consisted of 886 labeled and 7799 unlabeled abdominal CT 
        slices at the L3 vertebral level. The unlabeled images were used in a prior study exploring the impact of body composition on cardiovascular outcomes \cite{chaves2021opportunistic}.
        The labeled slices included manual segmentations for subcutaneous, visceral, and intramuscular adipose 
        tissue and muscle. These labeled slices were split into 
        709 slices for training, 133 for validation, and 44 for testing. 
        The training set was split in a similar manner as the MRI volumes into 4 additional subsets of 
        50\% (354 slices), 25\% (177 slices), 10\% (71 slices), and 5\% (35 slices) 
        training data. No metadata from the dataset was used in any models.
        
        \begin{table}[h!]
            \centering
            \caption{Demographics of the subjects included in 
            this study. Age is shown as mean $\pm$ standard deviation. 
            For the CT dataset, one subject did not have age information and 
            four subjects did not have gender information. \newline}
            \begin{tabular}{c c c c}\toprule
                \multicolumn{4}{c}{\textbf{MRI}} \\
                \midrule
                \textbf{Split} & \textbf{Gender} & 
                \textbf{\# Volumes (\# Slices)} 
                & \textbf{Age (range)}\\
                \midrule
                \multirow{3}{*}{Train} & Male & 46 (7360) & 44.7 $\pm$ 17.7 (17 - 75) \\
                & Female & 40 (6400) & 42.9 $\pm$ 18.5 (16 - 87) \\
                & Total & 86 (13760) & 43.9 $\pm$ 18.1 (16 - 87) \\
                \midrule
                \multirow{3}{*}{Validation} & Male & 18 (2880) & 37.3 $\pm$ 16.8 (18 - 68) \\
                & Female & 15 (2400) & 53.2 $\pm$ 14.9 (18 - 79) \\
                & Total & 33 (5280) & 44.5 $\pm$ 17.8 (18 - 79) \\
                \midrule
                \multirow{3}{*}{Test} & Male & 26 (4156) & 37.9 $\pm$ 14.9 (18 - 71) \\
                & Female & 10 (1584) & 53.0 $\pm$ 11.9 (31 - 73) \\
                & Total & 36 (5740) & 42.1 $\pm$ 15.6 (18 - 73) \\
                \midrule
                \multirow{3}{*}{Unlabeled} & Male & 37 (5446) & 38.1 $\pm$ 16.9 (15 - 77) \\
                & Female & 49 (6686) & 52.1 $\pm$ 18.5 (14 - 97) \\
                & Total & 86 (12132) & 46.1 $\pm$ 19.1 (14 - 97) \\
                \midrule
                \multicolumn{4}{c}{\textbf{CT}} \\
                \midrule
                \textbf{Split} & \textbf{Gender} & \textbf{\# Slices} 
                & \textbf{Age (range)}\\
                \midrule
                \multirow{3}{*}{Train} & Male & 362 & 68.2 $\pm$ 11.4 (20 - 97) \\
                & Female & 343 & 71.1 $\pm$ 10.5 (18 - 95) \\
                & Total & 709 & 69.6 $\pm$ 11.1 (18 - 97) \\
                \midrule
                \multirow{3}{*}{Validation} & Male & 63 & 69.1 $\pm$ 9.5 (32 - 83) \\
                & Female & 69 & 71.0 $\pm$ 11.0 (32 - 89) \\
                & Total & 133 & 70.1 $\pm$ 10.4 (32 - 89) \\
                \midrule
                \multirow{3}{*}{Test} & Male & 18 & 70.6 $\pm$ 11.9 (47 - 92) \\
                & Female & 26 & 73.1 $\pm$ 11.7 (44 - 93) \\
                & Total & 44 & 72.1 $\pm$ 11.9 (44 - 93) \\
                \midrule
                \multirow{3}{*}{Unlabeled} & Male & 3167 & 51.5 $\pm$ 17.1 (18 - 101) \\
                & Female & 4632 & 51.6 $\pm$ 17.1 (18 - 100) \\
                & Total & 7799 & 51.6 $\pm$ 17.1 (18 - 101) \\
                \bottomrule
            \end{tabular} 
            \label{table:data_distribution}
        \end{table}
        
    \subsection{Data Preprocessing}
        All models segmented 2D slices for MRI and CT images. 
        Each CT image was preprocessed at 
        different windows and levels (W/L) of HU to emphasize different image contrasts, resulting in three-channel images: soft-tissue (W/L=400/50), bone (W/L=1800/40), and a custom setting (W/L=500/50). 
        All images were normalized to have zero mean and unit standard deviation, with MR images normalized by volume and CT images normalized per channel. 
        
        \begin{figure}[h!]
            \centering 
            \includegraphics[width=\textwidth, height=3.5in]{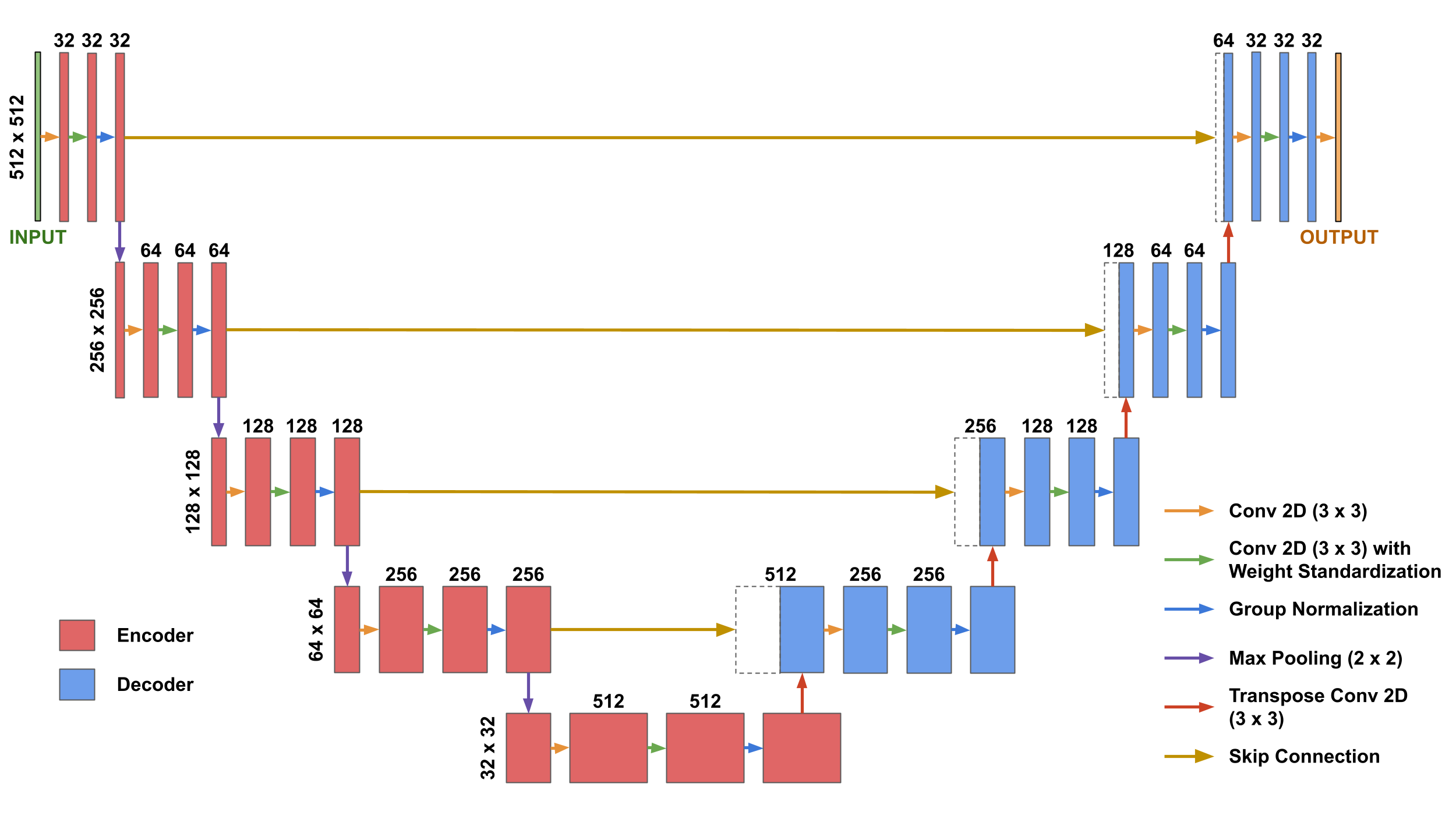}          
            \caption{The U-Net architecture used for both inpainting and segmentation, which includes layers grouped into three categories: the ``encoder" (in red), the ``decoder" (in blue), and the ``post-processing" layer (the final convolutional layer). 
            Each dotted rectangular box represents
            a feature map from the encoder that was concatenated to the first feature map in the decoder at 
            the same level.}
            \label{fig:unet_figure}
        \end{figure}
        
    \subsection{Model Architecture and Optimization}
        2D U-Net models \cite{ronneberger2015u} with Group Normalization 
        \cite{wu2018group}, weight standardization \cite{qiao2019micro}, 
        and He random weight initializations \cite{he2015delving} were used for inpainting and segmentation (Figure \ref{fig:unet_figure}). 
        Both inpainting and segmentation used identical U-Nets, except for
        the final convolutional layer, which we refer to as the ``post-processing" layer. For inpainting,
        the post-processing layer produced an output image with the same number of channels as the input image, whereas for segmentation, it produced a 4-channel image for the four 
        segmentation classes in each dataset. 
        
        We used L2 norm loss for inpainting and Dice loss, aggregated over 
        mini-batches per segmentation class, for segmentation.
        All training was performed with early stopping and the ADAM optimizer \cite{kingma2014adam} ($\beta_1$=0.99 and $\beta_2$=0.995) with a batch size of 9 on an NVIDIA 2080Ti GPU.
        Additional details are in the Supplementary Material 
        (Section \ref{sec:training_details}).
        
    \subsection{Image Corruption for Pretext Tasks}
        
        We incorporated random block selection to select square image patches during pretraining. 
        For context prediction, we set randomly selected patches of dimensions \emph{K}x\emph{K}
        to 
        zero until they formed at least 1/4th of the total image area.
        For context restoration, randomly selected pairs of 
        non-overlapping \emph{K}x\emph{K} image patches were swapped in an iterative manner
        until the number of corrupted pixels was at least 
        1/4th the total image area. 
        We refer to the result of both methods as ``masks". 
        The context prediction binary mask specified which pixels were zero and the context restoration mask was a list of patch pairs
        to be swapped.
        When pretraining with multi-channel CT images, 
        the locations of the 
        patch corruptions were identical across channels to avoid 
        shortcut learning \cite{geirhos2020shortcut}. 
        Example image corruptions are shown in Figure \ref{fig:example_images}. 
        
        To train the model to inpaint any arbitrarily corrupted image region without memorization of 
        image content, we sampled a random mask
        every iteration for all images. 
        For computational efficiency, we
        precomputed 100 random masks before training. We further randomly rotated the masks by either 0, 90, 180, or 
        270$^{\circ}$ counter-clockwise to increase the effective number of masks used during training to 400.
        
    \subsection{Design Choices for SSL Implementation}
    \label{sec:all_design_choices}
    
    Design choices for inpainting-based SSL segmentation revolving around pretraining task implementations \cite{pathak2016context, chen2019self} and transfer learning \cite{kornblith2019better, newell2020useful, park2019deep} have not been systematically compared. To overcome these shortcomings, we explored the 
    following questions: 
    
    \begin{enumerate}
        \item Which pretrained weights should be transferred for fine-tuning?
        \item How should the transferred pretrained weights be fine-tuned?
        \item What should be the initial learning rate when fine-tuning?
        \item What patch size should be used when corrupting images for inpainting?
        \item How should the locations of the patches be sampled when corrupting images for inpainting?
    \end{enumerate}
    
        \subsubsection{Design Choices for Transfer Learning (\#1-3)}
        \label{sec:design_choice_transfer}
        
        For design choice \#1 (which pretrained weights to transfer), we 
        compared transferring only the U-Net encoder weights \cite{pathak2016context} 
        with transferring both the encoder and decoder weights \cite{chen2019self}. 
        
        For design choice \#2, 
        we compare 
        (i) fine-tuning all pretrained weights 
        immediately after transferring \cite{kornblith2019better, newell2020useful}, and
        (ii) freezing pretrained weights after transferring and training 
        until convergence, then subsequently unfreezing pretrained 
        weights and training all weights until convergence
        \cite{park2019deep, kumar2022fine}.
        
        For design choice \#3,
        we selected four initial learning rates: 1e-2, 1e-3, 1e-4, and 1e-5, to evaluate whether pretrained features are distorted with larger learning rates. 
        
        To compare different combinations of these three design choices, we performed a grid 
        search and defined the best combination to be the one with
        the best
        segmentation performance on the MRI test set when trained with the MRI training subset 
        with 5\% training data. More details are in the Supplementary 
        Material (Section \ref{sec:sup_design_transfer_methods}).
        
        \subsubsection{Design Choices for Pretraining (\#4-5)}
        \label{sec:design_choice_pretraining}
        
        For design choice \#4, we compare patch sizes of 64x64, 32x32, 16x16, and 8x8 (Figure \ref{fig:example_images}).
        
        For design choice \#5, we compare two sampling methods: i) 
        fully-random sampling where the location of each patch was 
        selected at random and constrained to lie completely within the image \cite{pathak2016context, chen2019self}, and (ii) Poisson-disc
        sampling that enforces the centers of all \emph{K}×\emph{K} patches to lie at least $K\sqrt{2}$ pixels away from each other to prevent overlapping patches \cite{bridson2007fast}.
        
        To compare different combinations of design choices \#4 and \#5 and the two 
        pretext tasks, we performed a grid search by training a model for each combination 
        five times, 
        each time using one of the five training data subsets, for both datasets.
        We also trained a fully supervised model for each dataset and training data subset
        for a baseline comparison. 
        
        All models 
        were fine-tuned in an identical manner with the same random seed 
        after pretraining, using the best combination of design choices \#1-3.
        All inpainting models were compared by computing the L2 norm
        of the generated inpainted images. 
        All segmentation models were compared by computing the Dice coefficient for each 
        segmentation class in the test set, averaged across all available volumes/slices.
        
        \subsubsection{Optimal Pretraining Evaluation}
        \label{sec:choosing_best_ssl}
        
        We defined the optimal pretraining strategy as the strategy that provided the most benefit
        over supervised learning, across image modalities and training data extents, in the 
        experiment described in Section \ref{sec:design_choice_pretraining}. 
        
        For each baseline (fully supervised model) and SSL model trained in the 
        experiment using 50\%, 25\%, 10\%, and 5\% training data,
        we computed class-averaged Dice scores for every test volume/slice in 
        the MRI and CT datasets. 
        For each pretraining strategy and dataset,
        we compared whether the set of 
        Dice scores of the corresponding SSL models were significantly higher than that 
        of the respective fully-supervised models using one-sided Wilcoxon signed-rank tests.
        As a heuristic, the pretraining strategies were sorted by their associated p-values and 
        the pretraining strategy that 
        appeared in the top
        three for both the MRI and CT datasets was selected as the 
        optimal pretraining strategy. We defined
        the optimally trained model for each dataset as the SSL model that was 
        pretrained with this optimal pretraining strategy and fine-tuned for 
        segmentation using the best combination of design choices \#1-3.
        
    \subsection{Impact of Extent of Unlabeled Data}
    \label{sec:unlabeled_data_extent}
    
    To measure the effect of the number of pretraining images on downstream 
    segmentation performance, 
    the optimally trained model was pretrained with the standard training 
    set 
    as well as two supersets of the training set containing
    additional unlabeled imaging data. 
    We refer to the standard training set as 100\% pretraining data 
    (86 volumes for MRI and 709 slices for CT).
    For the MRI dataset, the second and third sets consisted of 150\% (129 volumes) and 
    200\% (172 volumes) pretraining data, respectively. For the CT dataset, the second and 
    third sets consisted of 650\% (4608 slices) and 1200\% (8508 slices) pretraining data, respectively.
    After pretraining, all the pretrained models were fine-tuned with the 
    five subsets of labeled training data and a Dice score was computed for each fine-tuned model, averaged across all segmentation classes and all volumes/slices in the test set. 
    
    For MRI and CT, the pretraining dataset that led to the best average Dice score 
    across the extents of labeled training data was chosen for further experiments.
    
    \subsection{Comparing SSL and Fully-Supervised Learning}
    
    We compared baseline fully-supervised models and the optimally trained models 
    pretrained with the chosen pretraining dataset from 
    the experiment described in Section \ref{sec:unlabeled_data_extent}.
    For each training data subset, models were evaluated using two clinically-relevant metrics for determining cartilage, muscle, and adipose tissue health status.
    For MRI, we computed mean T2 relaxation time
    per tissue and tissue volume \cite{Sveinsson2017}.
    For CT, we computed cross-sectional area and mean HU value
    per tissue. We calculated their percentage errors by comparing them to values 
    derived from using ground truth segmentations to compute the metrics.
        
    \subsection{Statistical Analysis}
    
    All statistical comparisons
    were computed using one-sided Wilcoxon signed-rank tests. All statistical 
    analyses were performed using the SciPy (v1.5.2) library \cite{virtanen2020scipy}, with Type-1 $\alpha=0.05$.

\section{Results}

    The subject demographics of all labeled and unlabeled 
    volumes/slices
    are shown in Table \ref{table:data_distribution}.

    \subsection{Design Choices for Transfer Learning}
    
        We observed that all pretrained model variants
        had high performance
        when first fine-tuned with an initial learning rate 
        of 1e-3 and then fine-tuned a second time with an initial learning rate of
        1e-4. Transferring pretrained encoder weights only and 
        fine-tuning once immediately with an 
        initial learning rate of 1e-3 achieved similar performance, with the 
        added benefit of reduced training time. Consequently, we used 
        these as the best combination of the three design choices 
        for transfer learning. 
        Additional details are in the Supplementary 
        Material (Section \ref{sup_design_transfer_results}).
        
    \subsection{Design Choices for Pretraining}
    \label{sec:results_design_choice_pretrain}
    
        The L2 norm 
        consistently decreased as a function of patch size for all combinations of 
        pretext tasks (context prediction and context restoration) and 
        sampling methods (random 
        and Poisson-disc) (Table \ref{table:inpainting_metrics}). Furthermore, 
        L2 norms for Poisson-disc sampling were 
        significantly lower than those for random 
        sampling (p$<$0.05).
        For all combinations,
        inpainted MRI images were of higher quality than inpainted CT images.
        
\begin{figure}[h]
            \centering 
            \includegraphics[width=\textwidth]{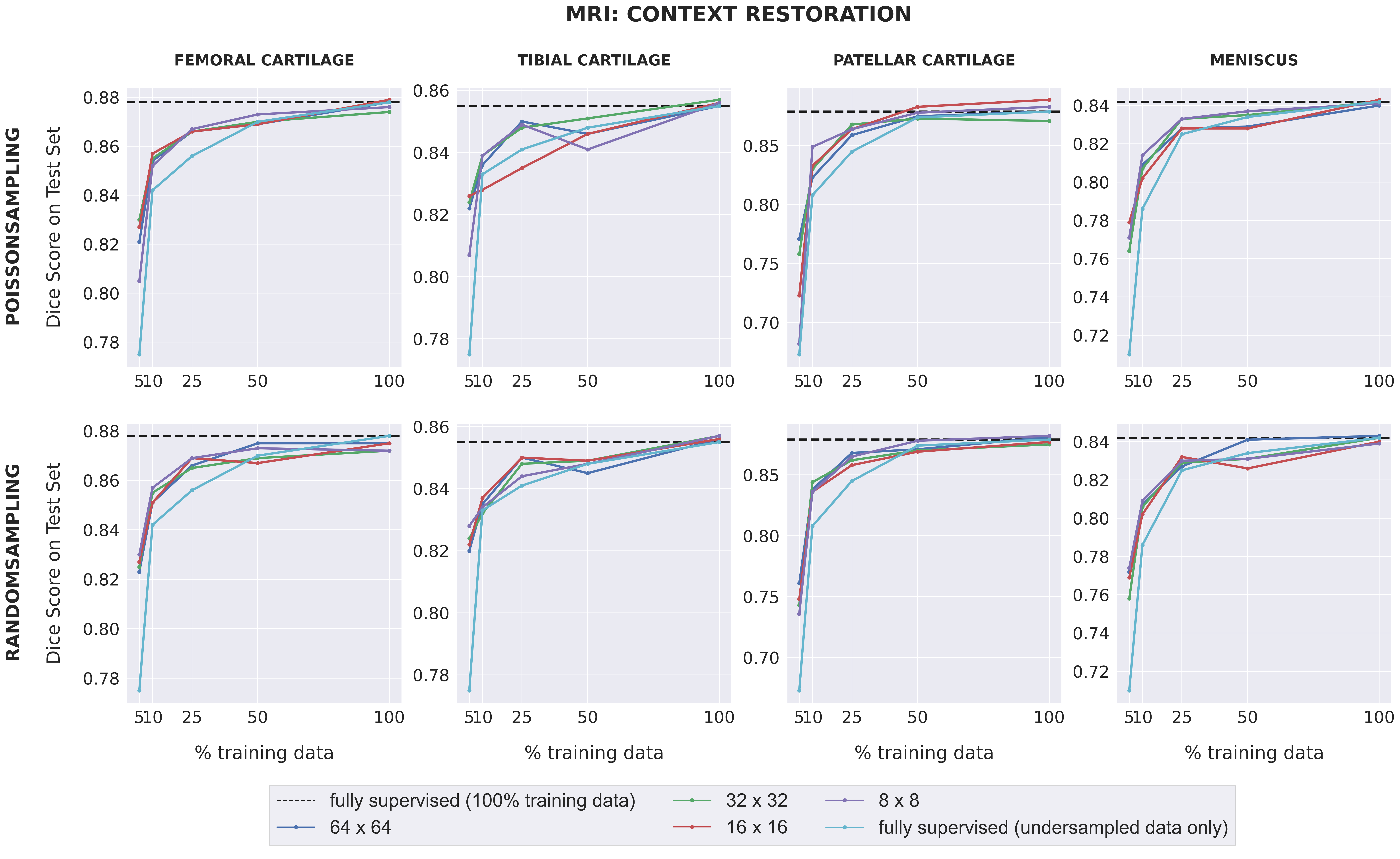}
            \caption{The downstream segmentation performance on the MRI dataset for the Context Restoration pretext task as measured by 
            the Dice score 
            for every combination of 
            patch size and sampling method used during pretraining, evaluated 
            in five different scenarios of training data availability. 
            In each scenario, every model is
            trained for segmentation using one of the 
            five different subsets 
            of training data as described in Section \ref{sec:mri_dataset}. 
            The black dotted line
            in each plot indicates the performance 
            of a fully supervised model trained using all available training images. 
            The light blue 
            curve indicates the performance of a fully supervised model when trained using 
            each of the five 
            different subsets of training data. Similar plots for the Context Prediction pretext task are given in the Supplementary Section.}
            \label{fig:mri_patch_size_figure}
        \end{figure}

        Dice scores for fully supervised baselines ranged from 
        0.67-0.88 across subsets of training data for MRI images. 
        Downstream segmentation performance for 
        the MRI dataset was similar for 
        all combinations of pretext task, 
        patch size, and sampling method (Figure \ref{fig:mri_patch_size_figure}). All SSL
        models matched (within 0.01) or outperformed the fully supervised model 
        in low-label regimes with 25\% training data or less for the femoral cartilage, patellar cartilage, 
        and meniscus, and had comparable performance for 
        higher data extents. For the tibial cartilage, all SSL models outperformed the fully 
        supervised model when trained on 5\% training data and had comparable performance for 
        higher data extents. The difference in Dice score between 
        each self-supervised model and the fully supervised model generally increased as 
        the amount 
        of labeled training data decreased.
        SSL pretraining also enabled some models to outperform the 
        fully supervised model trained with 100\% training data in 
        patellar cartilage segmentation.
        
\begin{figure}[h]
            \centering 
            \includegraphics[width=\textwidth]{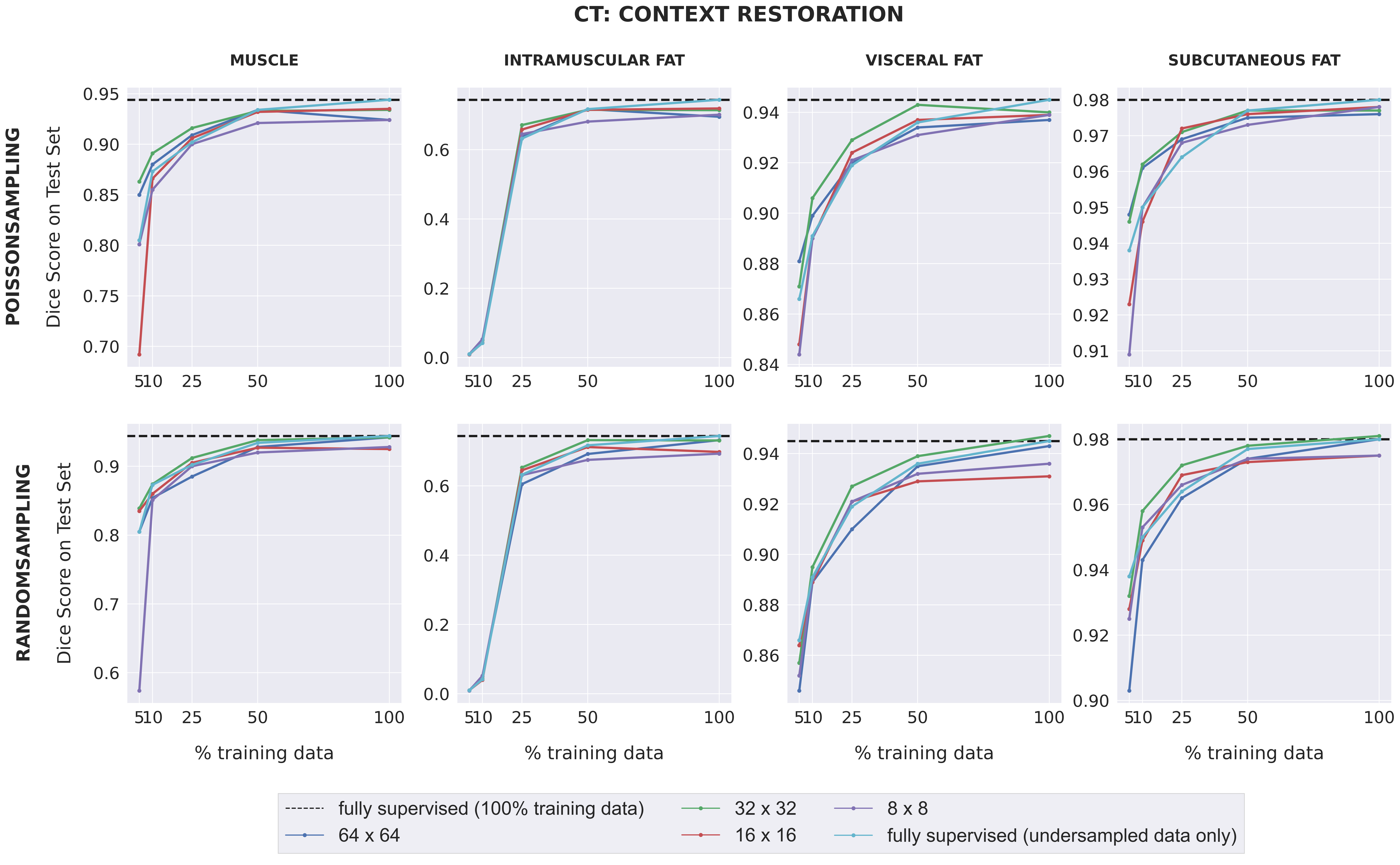}
            \caption{The downstream segmentation performance on the CT dataset for the Context Restoration pretext task as measured by 
            the Dice score 
            for every combination of 
            patch size and sampling method used during pretraining, 
            evaluated in five different scenarios of training data availability. 
            In each scenario,
            every model is trained for segmentation using one of the 
            five different subsets 
            of training data as described in Section \ref{sec:ct_dataset}. 
            The black dotted line 
            in each plot indicates the performance 
            of a fully supervised model trained using all available training images. 
            The light blue 
            curve indicates the performance of a fully supervised model when trained using 
            each of the five 
            different subsets of training data. Similar plots for the Context Prediction pretext task are given in the Supplementary Section.}
            \label{fig:ct_patch_size_figure}
        \end{figure}

        Dice scores for fully supervised baselines were consistently higher for CT images
        than for MRI images, with the exception of intramuscular adipose tissue.
        Unlike with the MRI dataset, downstream SSL segmentation 
        for CT in low-label regimes depended on the 
        pretext task and 
        the patch size used during pretraining 
        (Figure \ref{fig:ct_patch_size_figure}). Models pretrained with
        larger patch sizes (64x64; 32x32) often outperformed those 
        pretrained with smaller patch sizes (16x16; 8x8) for muscle, 
        visceral fat, and subcutaneous fat segmentation, when trained with 
        either 5\% or 10\% labeled data.
        Furthermore, when 25\%
        training data or less was used, models pretrained
        with 32x32 patches using context restoration almost always outperformed fully 
        supervised models for muscle, visceral fat, and subcutaneous fat 
        segmentation, but 
        rarely did so when pretrained using context prediction.
        For intramuscular fat, all SSL models had
        comparable performance with fully supervised models in 
        low-label regimes. For high-label regimes (over 25\% labeled data), 
        all SSL models had comparable performance with fully supervised 
        models for all four segmentation classes.
        
        \begin{table}[h]
            \centering
            \caption{Quantitative evaluation of inpainting for every combination of pretext task, patch size, and 
            sampling method. All values are rounded to the nearest integer.\newline}
            \resizebox{\textwidth}{!}{
                \begin{tabular}{c c c c c c}
                    \toprule
                    \multicolumn{2}{c}{\multirow{2}{*}{\textbf{Pretext Task and Patch Size}}} &
                    \multicolumn{2}{c}{\textbf{MRI}} &
                    \multicolumn{2}{c}{\textbf{CT}} \\
                    \cmidrule(lr){3-4}
                    \cmidrule(lr){5-6}
                    \multicolumn{2}{c}{}
                    & \multicolumn{2}{c}{\thead{\textbf{L2 Norm} \\ \textbf{(mean $\pm$ std)}}}
                    & \multicolumn{2}{c}{\thead{\textbf{L2 Norm} \\ \textbf{(mean $\pm$ std)}}} \\
                    \midrule
                    \textbf{Pretext Task} & \textbf{Patch Size} & \textbf{Poisson-Disc} & \textbf{Random} & 
                    \textbf{Poisson-Disc} & \textbf{Random} 
                    \\
                    \midrule
                    \multirow{4}{4em}{\thead{\textbf{Context} \\ \textbf{Prediction}}} 
                    & \textbf{64x64} & 
                    94 $\pm$ 9 &
                    105 $\pm$ 9 & 
                    123 $\pm$ 13 &
                    134 $\pm$ 18 \\
                    & \textbf{32x32} & 
                    75 $\pm$ 8 &
                    81 $\pm$ 8 &
                    83 $\pm$ 9 & 
                    112 $\pm$ 14 \\
                    & \textbf{16x16} & 
                    61 $\pm$ 7 &
                    64 $\pm$ 7 &
                    66 $\pm$ 8 &
                    74 $\pm$ 10 \\
                    & \textbf{8x8} & 
                    51 $\pm$ 6 &
                    52 $\pm$ 5 &
                    54 $\pm$ 7 &
                    57 $\pm$ 8 \\
                    \midrule
                    \multirow{4}{4em}{\thead{\textbf{Context} \\ \textbf{Restoration}}} 
                    & \textbf{64x64} & 
                    96 $\pm$ 9 &
                    116 $\pm$ 11 &
                    142 $\pm$ 19 &
                    346 $\pm$ 158 \\
                    & \textbf{32x32} & 
                    75 $\pm$ 8 &
                    84 $\pm$ 8 &
                    108 $\pm$ 12 &
                    127 $\pm$ 15 \\
                    & \textbf{16x16} & 
                    62 $\pm$ 7 &
                    67 $\pm$ 7 &
                    86 $\pm$ 10 &
                    93 $\pm$ 12 \\
                    & \textbf{8x8} & 
                    51 $\pm$ 5 &
                    56 $\pm$ 6 &
                    66 $\pm$ 8 &
                    80 $\pm$ 9 \\
                    \bottomrule
                \end{tabular}
            }
            \label{table:inpainting_metrics}
        \end{table}         
        
    \subsection{Optimal Pretraining Evaluation}
    
        The top 5 pretraining strategies for the MRI dataset and the top 3 pretraining 
        strategies for the CT dataset led to significantly better segmentation performance
        compared to fully supervised learning
        (p$<$0.001) (Table \ref{table:best_ssl_p_value}). 
        
        For MRI, the top 5 
        strategies all consisted of pretraining with context restoration, 
        with minimal differences in p-value based on 
        the patch size 
        and sampling method used. For CT, the top 5  strategies 
        used a patch size of at least 32x32 during pretraining.
        The strategy
        of pretraining with
        context restoration, 32x32 patches, and Poisson-disc sampling was in the top 
        3 for both datasets, and was therefore selected as the optimal 
        pretraining strategy.
        
        \begin{table}[h]
            \centering
            \caption{Summary of the top five combinations of pretext tasks, patch sizes, 
            and sampling methods for each 
            dataset with the corresponding p-value for each combination, and sorted by p-value in ascending order. 
            The bolded pretext task, patch size, and sampling method were chosen as the best combination of the three
            design choices. \newline}
            \resizebox{\textwidth}{!}{
                \begin{tabular}{c c c c c}
                    \toprule
                    \multicolumn{5}{c}{\textbf{MRI}} \\
                    \midrule
                    \textbf{Rank}
                    & \textbf{Pretext Task} & \textbf{Patch Size} & \textbf{Sampling Method} & \textbf{p-value} \\
                    \midrule
                    1 & Context Restoration & 64x64 & Random & 1.64 x $10^{-18}$ \\ 
                    2 & Context Restoration & 8x8 & Random & 1.89 x $10^{-18}$ \\
                    3 & \textbf{Context Restoration} & \textbf{32x32} & \textbf{Poisson-Disc} & 1.05 x $10^{-17}$ \\
                    4 & Context Restoration & 8x8 & Poisson-Disc & 4.03 x $10^{-17}$ \\
                    5 & Context Restoration & 32x32 & Random & 9.38 x $10^{-16}$ \\
                    \midrule
                    \multicolumn{5}{c}{\textbf{CT}} \\
                    \midrule
                    \textbf{Rank} 
                    & \textbf{Pretext Task} & \textbf{Patch Size} & \textbf{Sampling Method} & \textbf{p-value} \\
                    \midrule
                    1 & \textbf{Context Restoration} & \textbf{32x32} & \textbf{Poisson-Disc} & 6.29 x $10^{-17}$ \\
                    2 & Context Restoration & 64x64 & Poisson-Disc & 1.88 x $10^{-9}$ \\
                    3 & Context Restoration & 32x32 & Random & 4.12 x $10^{-7}$ \\
                    4 & Context Prediction & 64x64 & Poisson-Disc & 0.1 \\
                    5 & Context Prediction & 64x64 & Random & 0.66 \\
                    \bottomrule
                \end{tabular}
            }
            \label{table:best_ssl_p_value}
        \end{table}
        
    \subsection{Impact of Extent of Unlabeled Data}
    \label{sec:extent_unlabeled_results}
    
        For both datasets and for most subsets of labeled training data used during 
        fine-tuning (except 25\% and 10\% labeled training data for MRI), 
        the optimally trained
        model performed significantly better in downstream segmentation when 
        pretrained on the maximum amount of data per dataset 
        (200\% pretraining data 
        for MRI and 1200\% pretraining data for CT) than when pretrained on only the 
        training 
        set (p$<$0.05). When 25\% or 10\% labeled training data was used for MRI segmentation, 
        the optimally trained 
        model achieved a higher mean Dice score when pretrained on 200\% pretraining data,
        but this was not statistically significant (p=0.3 for 25\% labeled training data and 
        p=0.02 for 10\% labeled training data).
        
        For MRI, Dice scores 
        almost always improved as the amount of pretraining data 
        increased.
        This improvement was greatest when only 5\% of the labeled training data was 
        used for training segmentation. 
        Improvements in segmentation performance were slightly higher for CT.
        For all extents of labeled training data, segmentation performances 
        improved when the amount 
        of pretraining data increased from 
        100\% to 650\%. There was 
        limited improvement when the amount of pretraining data 
        increased from 650\% to 1200\%. 
        
        Pretraining on the maximum amount of data 
        enabled the 
        optimally trained models 
        to surpass 
        the performance of fully supervised models for all extents of labeled training data,
        in both MRI and CT 
        For the MRI dataset, the highest improvement over supervised learning was observed when 5\% labeled training data was used. For CT, 
        considerable improvements over supervised learning were observed 
        when 5\%, 10\%, or 25\%
        labeled training data was used.
        
        For both the MRI and CT datasets, the best average Dice score over 
        all extents of 
        labeled training data occurred when the maximum possible amount of 
        pretraining data was used (200\% pretraining data for MRI and 1200\%
        pretraining data for CT).
        
        \begin{figure}[h!]
            \centering 
            \includegraphics[width=\textwidth]{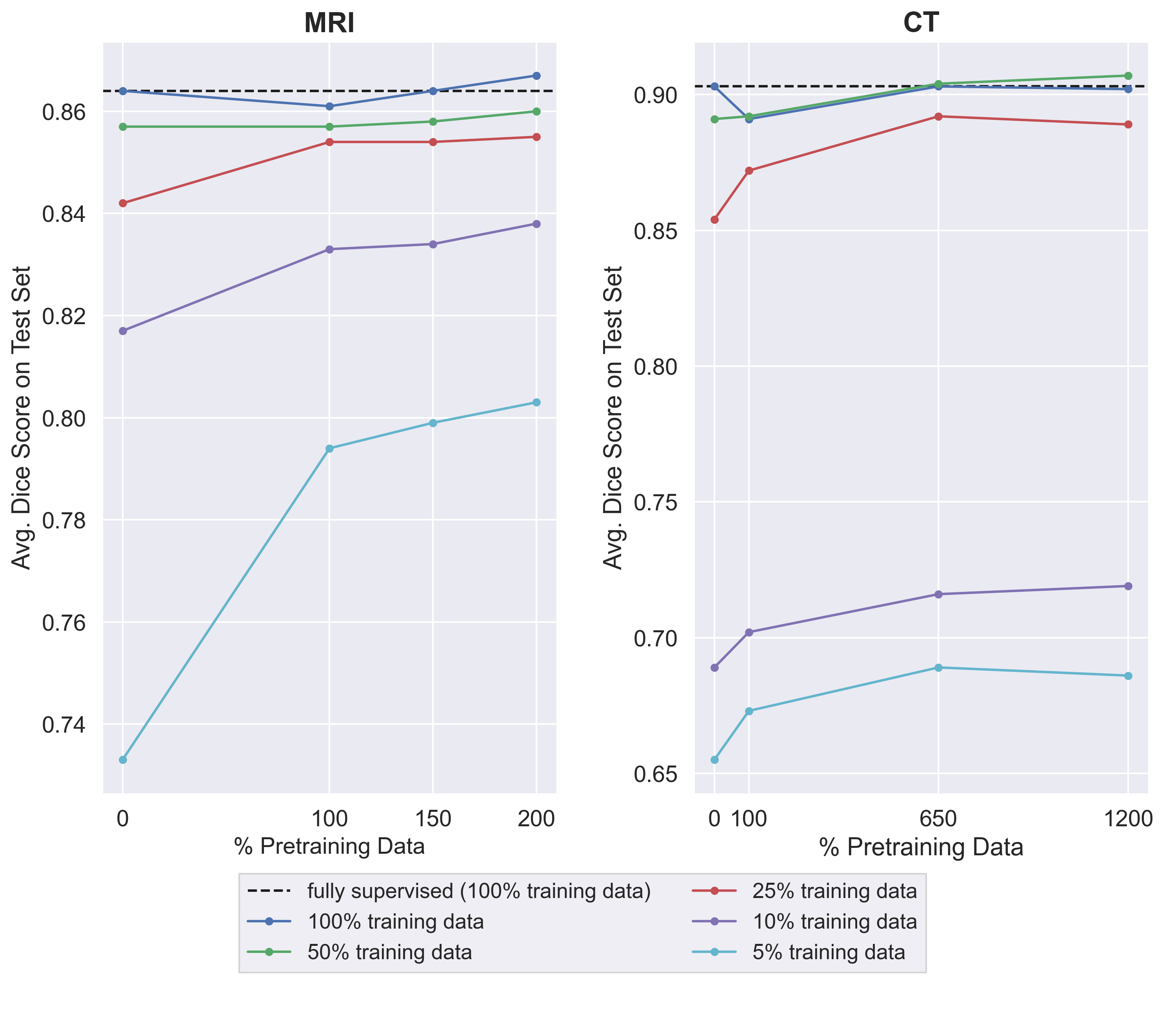}
            \caption{The downstream segmentation performance of the 
            optimally trained model when 
            pretrained with different amounts of pretraining data and fine-tuned using each 
            of the five training data subsets. 100\% pretraining data refers
            to the regular training set for each dataset. The data point for 0\% 
            pretraining data is the performance of a fully supervised model. The 
            black dotted line indicates the performance of a fully supervised model trained on all available training 
            data for the appropriate dataset.}
            \label{fig:num_pretrain_figure}
        \end{figure}
        
    \subsection{Comparing SSL and Fully-Supervised Learning}
    
        For each dataset, optimally trained models 
        were pretrained with the maximum amount of pretraining data from Section 
        \ref{sec:extent_unlabeled_results}.
         
\begin{figure}[h!]
            \centering 
            \includegraphics[width=\textwidth]{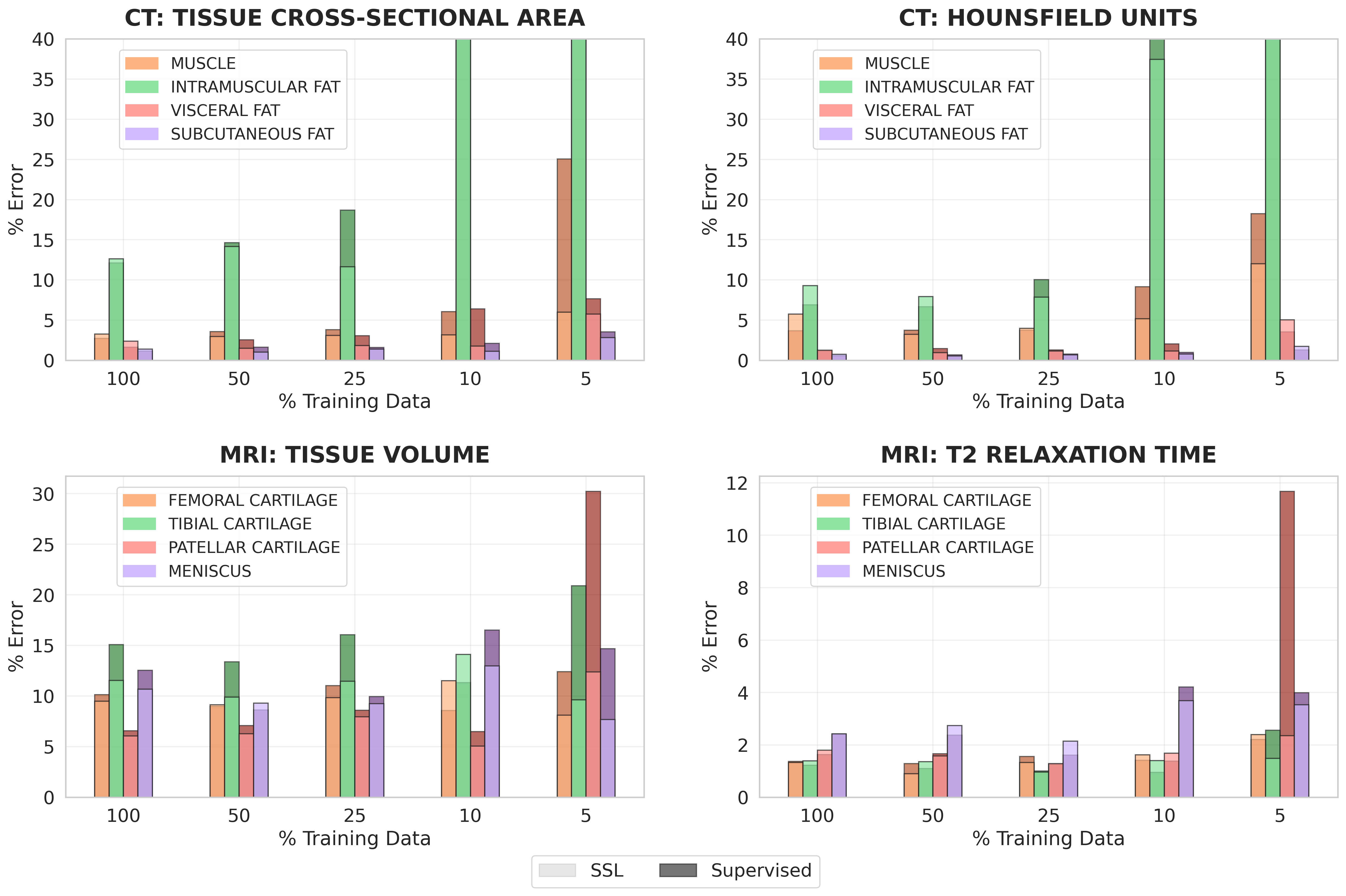}
            \caption{A comparison of the percent error in calculating 
            clinical metrics for the MRI and 
            CT datasets between when the tissue segmentations are generated by
            fully supervised models and when 
            the tissue segmentations are generated by
            optimally trained models, 
            pretrained using 200\% data for MRI and 1200\% data for CT. 
            Each bar represents the median percent error
            across the test set for a particular tissue, clinical metric, and label regime.
            The percent error
            in the calculation of tissue area and mean HU for intramuscular fat extends beyond the 
            limits of the y-axis when 10\% and
            5\% labeled training data for segmentation is used.}
            \label{fig:ssl_v_supervised_figure}
        \end{figure}

        For all clinical metrics, using  
        optimally trained models generally 
        led to lower percent errors than using
        fully supervised
        models in regimes of 10\% and 5\% labeled training data 
        (Figure \ref{fig:ssl_v_supervised_figure}).
        These differences were especially 
        pronounced for CT tissue area, MRI tissue volume, and MRI mean T2 
        relaxation time. With 5\% labeled training data for MRI,
        segmentations from optimally pertrained models more than halved 
        the percent error for both tissue volume and mean T2 relaxation 
        time of patellar cartilage, compared to segmentations 
        from fully supervised models. 
        
        With 100\% or 50\% labeled training data, 
        percent errors for all clinical metrics had lower improvement 
        when 
        optimally trained models were used. However, for MRI tissue volume, 
        which requires accurate 
        segmentation contours,
        optimally trained models almost always outperformed the 
        fully supervised models,
        even in scenarios with large amounts of 
        labeled training data.
        
        For both datasets, clinical metrics improved the most for the 
        most challenging classes 
        to segment. This included intramuscular adipose tissue for CT, where percent error 
        decreased from around 3940\% to 3600\% for tissue area when 10\% labeled training data was 
        used, and patellar cartilage for MRI, where percent error decreased from around 
        30\% to 12\% for tissue volume when 5\% labeled training data was used.
        
\section{Discussion}

        In this work, we investigated several key, yet under-explored design choices associated
        with pretraining and transfer learning in inpainting-based SSL for tissue segmentation.
        We examined the effect of inpainting-based SSL on the performance of tissue 
        segmentation in various data and label regimes for MRI and CT scans, 
        and compared it with fully supervised learning. We quantified performance
        using standard Dice scores
        and four clinically-relevant metrics of imaging biomarkers.   
        
        We observed that the crosstalk between the initial and fine-tuning learning rate was a design choice that most affected model performance. All model variants achieved optimal
        performance with an
        initial learning rate 
        of 1e-3 and a fine-tuning
        learning rate of 1e-4 (Figure \ref{fig:fine-tuning_figure}). 
        This suggests the need for not perturbing the pretrained representations 
        from the pretext task with a large learning rate.
        Moreover, although 
         freezing and then fine-tuning the transferred weights provided an improvement 
         over fine-tuning immediately for this learning rate combination 
         (Figure \ref{fig:fine-tuning_figure}), the improvement was very small. 
         This result matches 
         the findings of Kumar et al. \cite{kumar2022fine}, where the performance of linear probing (freezing) and 
         then fine-tuning only slightly improved the performance of fine-tuning 
         immediately after transferring. 
        Additional details are provided in the Supplementary Material (Section \ref{sup_design_transfer_discussion}). 
        
        Here, we suggest some best practices for inpainting-based SSL for medical imaging segmentation tasks. We observed that downstream segmentation performance for  
        MRI was similar for all combinations 
        of pretext tasks, 
        patch sizes, and sampling techniques.
        This observation remained consistent despite 
        significant differences in the
        L2 norms of the inpainted images.
        While decreasing patch sizes 
        and sampling patch locations via Poisson-disc sampling to ensure non-overlapping patches
        both resulted in significantly lower L2 norms, they did not improve downstream segmentation performance. 
        These observations suggest a discordance between learning semantically meaningful 
        representations and the accuracy of the pretext task metric. Thus, 
        simply performing \textit{good enough} pretraining may be more important 
        than optimizing pretext task performance. 
        
        For both MRI and CT, segmentation performance 
        usually increased in proportion to the amount of pretraining data.
        The highest improvements 
        over supervised learning were 
        observed in the context of very 
        low labeled data regimes of 5-25\% labeled 
        data.
        These empirical observations across both MRI and 
        CT demonstrate that pretraining 
        with large enough datasets improves 
        performance compared to only supervised training, 
        especially when the amount of available training data is limited. 
        
        Similar to supervised learning, improvements in 
        SSL Dice scores tended to follow a power-law relationship as the size of the unlabeled 
        corpora increased 
        \cite{desai2019technical}. 
        The observation that pretraining 
        on 650\% and 
        1200\% CT pretraining data led to similar improvements 
        over supervised learning suggests a limit
        exists where the learning capacity of a model 
        saturates and additional unlabeled data may 
        not improve downstream performance.
        A good practice for future segmentation studies may be to 
        create 
        Figure \ref{fig:num_pretrain_figure} 
        to evaluate the trade-off between the challenges of annotating more 
        images and acquiring more unlabeled images. 
       
        Compared to fully-supervised models, optimally trained 
        models generally 
        led to more accurate values for all clinical metrics in label-limited scenarios.
        We also observed that clinical metrics improved the most with SSL for tissue classes 
        that had the highest percent error with fully supervised learning - 
        intramuscular adipose tissue in CT and patellar cartilage 
        in MRI. 
        This observation, combined with the Dice score improvement in low labeled data regimes,
        suggests that SSL may be most efficacious when 
        the performance of 
        the baseline fully supervised model is low.
        
        When training with 5\% labeled data for all MRI classes and muscle on CT, our optimal pretraining strategy improved Dice scores by over 0.05, compared to fully supervised learning. 
        In such cases, the Dice score for fully supervised learning was 0.8 or lower, which suggests a critical performance threshold where inpainting-based SSL can improve segmentation performance over supervised learning.
        SSL may be beneficial in these cases 
        because the models still have the capacity 
        to learn more meaningful representations,
        compared to models with Dice scores over 0.8 that may already be saturated in their capacity to represent the underlying image. 
        
        Importantly, it should be noted that 
        the improvement in segmentation performance with SSL pretraining in 
        label-limited scenarios is 
        on the similar order 
        as prior advances that used complex DL architectures and training strategies 
        \cite{dai2021can3d, perslev2021cross, panfilov2019improving}. 
        Comparatively, our proposed SSL training paradigm offers an easy-to-use framework 
        for improving model performance without requiring large
        and difficult to train DL models. 
        
    \subsection{Study Limitations}
    
        There were a few limitations with this study. Although we investigated two 
        different methods for selecting which 
        pretrained weights to transfer, we did not conduct a systematic study across all 
        possible choices 
        due to computational constraints that made searching over the large 
        search space too inefficient. 
        We also leave other SSL strategies such as contrastive learning to future studies since it requires systematic evaluation of augmentations and sampling strategies.
        Furthermore, when we investigated the impact of unlabeled data extents 
        on downstream 
        segmentation performance, we did not pretrain our SSL models with 
        equal extents of unlabeled MRI and CT data since we maximized the 
        amount of available MRI data. 
        Finally, our investigations in this work are limited to 
        the U-Net architecture, though future work can explore other 
        powerful segmentation architectures.
        
\section{Conclusion}

In this work, we investigated how inpainting-based SSL improves MRI and CT segmentation compared to fully-supervised learning, especially in label-limited regimes. We presented an optimized training strategy and open-source implementation for performing such pretraining. We describe the impact of pretraining task optimization and the relationship between the sizes of labeled and unlabeled training datasets. Our proposed approach for pretraining for improving segmentation performance that does not require additional manual annotation, complex model architectures, or model training techniques. 

\bibliographystyle{IEEEtran}
\bibliography{IEEEabrv, references}

\clearpage

\section{Supplementary Material}
\label{sec:supplementary_material}

     \subsection{Implementation Details}
     
        \subsubsection{Model Architecture and Optimization}
        \label{sec:training_details}
        
        The loss function for inpainting was the L2 loss, implemented as in 
        Equation \ref{eq:L2} for a model 
        output ($X$) and ground truth ($Y$), where 
        $N$, $H$, $W$, and $C$  
        denote the batch size, height, width, and number of channels of the 
        images, respectively. 
        
        \begin{equation}
        \label{eq:L2}
            L_{inpainting}(X, Y) = \frac{1}{C}\sum_{c=0}^{C-1}\frac{1}{N}\sum_{n=0}^{N-1}\sum_{h=0}^{H-1}\sum_{w=0}^{W-1}
            \left(X_{n, h, w, c} - Y_{n, h, w, c}\right)^2
        \end{equation}
        
        The loss function for segmentation was a variant of the Dice loss, implemented as in Equation \ref{eq:dice} for 
        a model output ($X$) and ground truth ($Y$), 
        where $N$, $H$, $W$, and $C$  
        denote the batch size, height, width, and number of channels of the 
        images, respectively. 
        Due to the instability of pixel-wise losses for sparse classes, we used 
        a batch-aggregate Dice loss, 
        where the Dice loss was computed over the aggregate of a mini-batch per 
        segmentation class and the final loss was the mean Dice loss across 
        segmentation classes.
        
        \begin{equation}
        \label{eq:dice}
            L_{segmentation}(X, Y) = \frac{1}{C}\sum_{c=0}^{C-1}
            1 - \frac{2 * \sum_{n=0}^{N-1}\sum_{h=0}^{H-1}\sum_{w=0}^{W-1}
            \left(X_{n, h, w, c} * Y_{n, h, w, c}\right) + \epsilon}
            {\sum_{n=0}^{N-1}\sum_{h=0}^{H-1}\sum_{w=0}^{W-1}\left(X_{n, h, w, c} + 
            Y_{n, h, w, c}\right) + \epsilon}
        \end{equation}
        
        For inpainting, the learning rate was set to 1e-3 and decayed by a 
        factor of 0.9 every 2 epochs. To prevent overfitting, early 
        stopping \cite{prechelt1998early} based on the validation L2 loss was used 
        with a threshold of 50 and patience of 4 epochs. 
        For a baseline fully-supervised segmentation, the learning rate followed 
        the same schedule as inpainting. For self-supervised segmentation, 
        the fine-tuning learning rate was considered a design choice and is 
        described in Section \ref{sec:design_choice_transfer}.         
        For both fully-supervised and self-supervised segmentation, 
        early stopping based on the validation Dice loss was used to 
        prevent overfitting, with a threshold of 1e-3 and a patience of 10 
        epochs. All inpainting and segmentation models were trained until the 
        criteria for early stopping was achieved. 
        The same random seed was used for all experiments.
        
        All models were trained using the Keras (v2.1.6) software library \cite{chollet2015keras}, 
        with Tensorflow (v1.15.0) \cite{tensorflow2015-whitepaper} as the backend.
        
        \subsubsection{Design Choices for SSL Implementation}
        
        For all experiments described in Section \ref{sec:all_design_choices}, 
        the self-supervised models were pretrained on all of the 
        training data for the appropriate dataset. 
        
    \subsection{Design Choices for Transfer Learning}
    
        \subsubsection{Grid Search Implementation}
        \label{sec:sup_design_transfer_methods}
        
        As described in Section \ref{sec:design_choice_transfer}, we compared two methods for selecting
        which pretrained weights to transfer and two methods for fine-tuning the transferred pretrained weights. 
        To compare the four combinations of these two design 
        choices, we trained one model per combination. Since the first fine-tuning 
        method, in which the pretrained weights are fine-tuned immediately, involves one training run, and the second 
        fine-tuning method, in which the pretrained weights are first frozen, 
        involves two training runs, we chose to train the two
        models in which the 
        pretrained weights were fine-tuned immediately two times to ensure a fair comparison.
        
        To investigate the impact of the initial learning rate during fine-tuning, 
        we trained each of the four 
        models four times during the first training run, each time with one of the four possible initial 
        learning rates, 
        and then trained each of the sixteen trained models again four times, each time with one of the 
        four possible initial learning rates. 
        
        We selected the learning rates 1e-2, 1e-3, 1e-4, and 1e-5 for specific reasons. 
        1e-2 was selected as an example of a large learning rate, 
        to determine if fine-tuning with a large learning rate will destroy pretrained
        features. 1e-3 was selected as an example of a "common" learning rate, and 
        was used as the initial learning rate for all our other experiments. Finally, 1e-4 and 1e-5 were 
        selected arbitrarily as 
        examples of small learning rates. 
        
        All pretrained weights were derived from an inpainting 
        model that was trained with context prediction with 16x16 patches 
        and Poisson-disc sampling, and all
        models were fine-tuned for segmentation using the MRI training subset with 5\% data. The same random 
        seed was used when training each model. All models were 
        compared by computing the Dice coefficient for each volume in the MRI test set, averaged
        across the four segmentation classes.
        
        \subsubsection{Results}
        \label{sup_design_transfer_results}
        
        \begin{figure}[h!]
        \centering 
        \includegraphics[width=\textwidth]{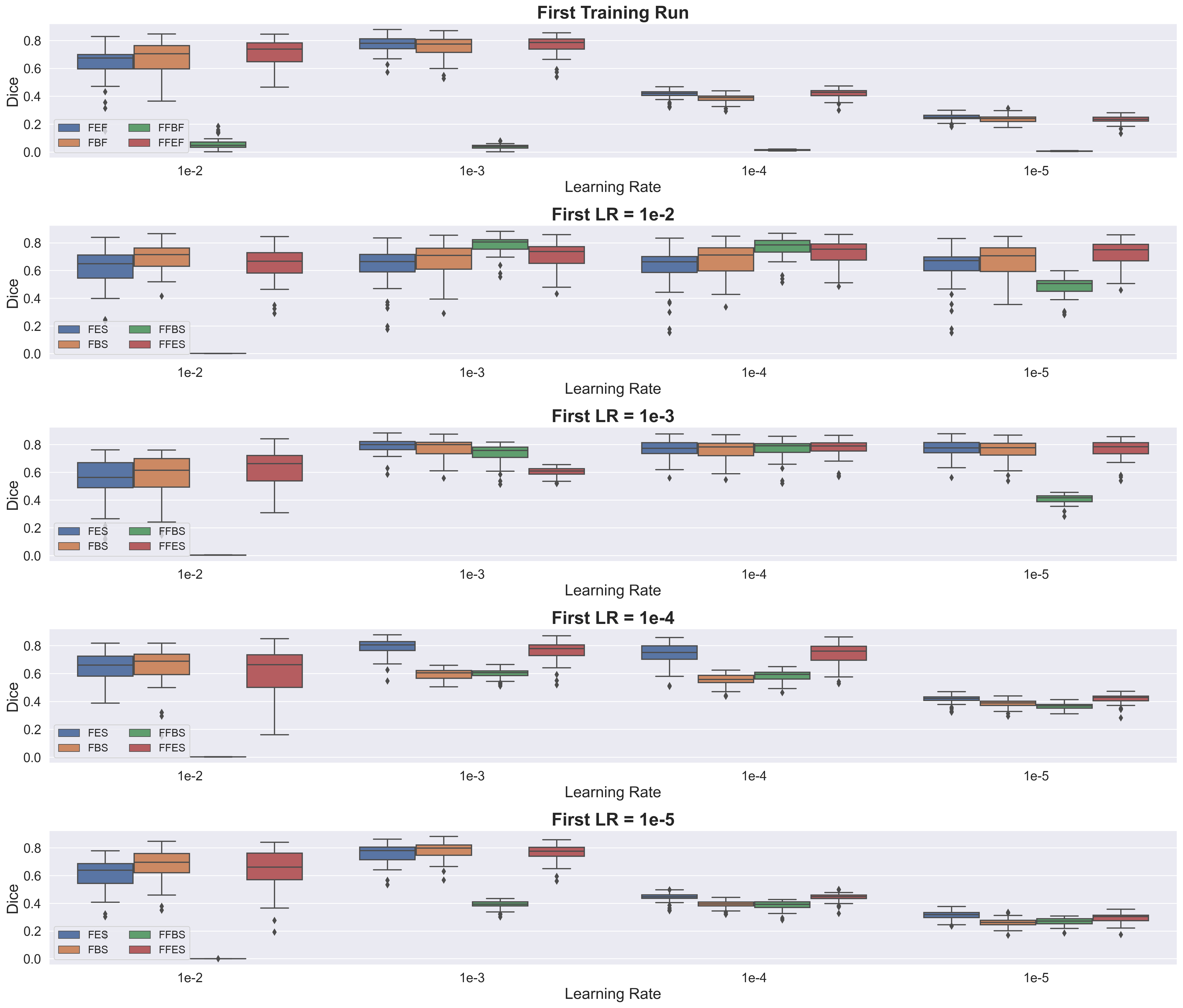}
        \caption{Box plots displaying the spread of Dice scores among the volumes in the MRI test set. The
        top row displays the spread of Dice scores after each type of model was trained once, 
        with the initial learning rate set to the appropriate value on the x-axis. 
        The remaining four rows display the spread of 
        Dice scores after each model in the first row was trained again, with the initial learning rate set 
        to the appropriate value on the x-axis. We used the following structure for acronyms to distinguish 
        between the different types of models: \textbf{ABC}. If \textbf{A} is F, the pretrained weights were
        fine-tuned immediately, and if \textbf{A} is FF, the pretrained weights were first frozen and then 
        fine-tuned. If \textbf{B} is E, only the pretrained encoder weights were transferred, and if 
        \textbf{B} is B, both the pretrained encoder and decoder weights were transferred. If \textbf{C} is 
        F, the model was trained only once (the first training run), and if \textbf{C} is S, the model was
        trained a second time (the second training run).}
        \label{fig:fine-tuning_figure}
        \end{figure}
    
        For the first training run following pretraining, higher initial learning rates of 1e-2 and 1e-3 
        produced better results. The FEF, FBF, and FFEF models had similar performance for all initial 
        learning rates, and consistently performed better than the FFBF models (Figure \ref{fig:fine-tuning_figure}).
         
        When each type of model was trained for a second time with an initial learning rate of 1e-2, 
        each model's performance was similar to its performance when trained only once with 
        an initial learning of 1e-2. This occurred regardless of the initial learning rate during the 
        first training run. For example, the FEF, FBF, and FFEF models had relatively high performance when 
        trained once with an initial learning rate of 1e-3, but when these models were trained for a 
        second time with an initial learning rate of 1e-2, the performance of all three models dropped to the 
        same level of performance as when each type of model was trained only once with an initial learning rate 
        of 1e-2. Similarly, but in the opposite way, the FEF, FBF, and FFEF models had relatively low 
        performance when trained once with an initial learning rate of 1e-4 or 1e-5, but when these models were 
        trained for a second time with an initial learning rate of 1e-2, the performance of all three
        models increased to the same level of performance as when each type of model was trained only once 
        with an initial learning rate of 1e-2.
         
        When each type of model was trained once with an initial learning rate of 1e-2 and then trained a 
        second time with a smaller learning rate, the FFBS models outperformed the FES and FBS models when the 
        initial learning rate of the second training run was 1e-3 or 1e-4, and the FFES models outperformed 
        the FES and FBS models for all initial learning rates smaller than 1e-2.
         
        When each type of model was trained once with an initial learning rate of 1e-4 and then trained a second 
        time with an initial learning rate of 1e-3 or lower, the FES and FFES models had similar performance 
        and always outperformed the FBS and FFBS models. 
         
        When each type of model was trained once with an initial learning rate of 1e-3, training each model 
        again with an initial learning rate equal to or less than 1e-3 did not improve or only slightly improved 
        the model's performance. The exception was the FFBF model, for which the performance 
        always increased by a large amount during the second training run, regardless of the 
        initial learning rate used during the second training run. The FEF, FES, FBF, and FBS models 
        had similar performance when the initial learning rates for the first 
        and second training runs were set to 1e-3.
         
        We concluded that FEF (fine-tuning immediately after transferring only the pretrained encoder 
        weights), 
        trained with an initial learning rate of 1e-3, was the best combination of design choices 
        for transfer learning because this model achieved high segmentation performance with minimal 
        training time.
         
        \subsubsection{Discussion}
        \label{sup_design_transfer_discussion}
        
        In this experiment, we determined the best combination of fine-tuning mechanism, 
        weight loading strategy, and initial learning rate during fine-tuning. Overall, every model had high performance when first trained with an initial learning 
        rate of 1e-3 and then trained a second time with an initial learning rate of 1e-4, despite using 
        different fine-tuning mechanisms and different methods for selecting which pretrained weights to transfer.
        This suggests choosing the initial learning rates for the 
        first and second training runs is the design choice for 
        transfer learning that most affects model performance. 
        
        If the initial learning rate of the first training run is 
        too large, like 1e-2, the pretrained features are at risk of being 
        destroyed. For example, the FEF and FBF models performed worse when trained
        with an initial learning rate of 1e-2 than when trained with an initial learning
        rate of 1e-3.
        Furthermore, when the initial learning rate during the second training run is too large, 
        a model has a risk of escaping out 
        of an 
        already found local minimum. For example, although the 
        FBF model had high performance when 
        trained once with 
        an initial learning rate of 1e-3, its performance dropped when trained again with an initial learning 
        rate of 1e-2.
        
        On the other hand, if the initial learning rate is too small, a model may not be able to learn during 
        fine-tuning. This was suggested by the low performance of all four types of 
        models when they were trained once with an initial learning rate of 1e-5 and then trained again 
        with an initial learning rate of either 1e-4 or 1e-5.
        
        Although the design choice for transfer learning that most affects model performance is the initial learning rate 
        during fine-tuning, the results of this experiment suggest that transferring only the 
        pretrained encoder weights 
        may lead to better performance gains than transferring both the pretrained encoder and decoder 
        weights.
        For instance, in the first training run, the FFEF models performed similarly to the FEF models 
        for all initial learning rates, suggesting the frozen encoder features in the FFEF models were as good 
        as the fine-tuned encoder features in the FEF models. In addition, when the four types of models were 
        first trained with an initial learning rate of 1e-4 and then trained again with an initial learning 
        rate of 1e-3 or lower, the FES and FFES models always outperformed the FBS and FFBS models. 
        These results suggest that transferring only the pretrained encoder weights provides a 
        better initialization
        point for segmentation fine-tuning than transferring both the pretrained encoder and decoder weights.
        
    \subsection{Design Choices for Pretraining: Supplementary Figures}
    
        \begin{figure}[h]
            \centering 
            \includegraphics[width=\textwidth]{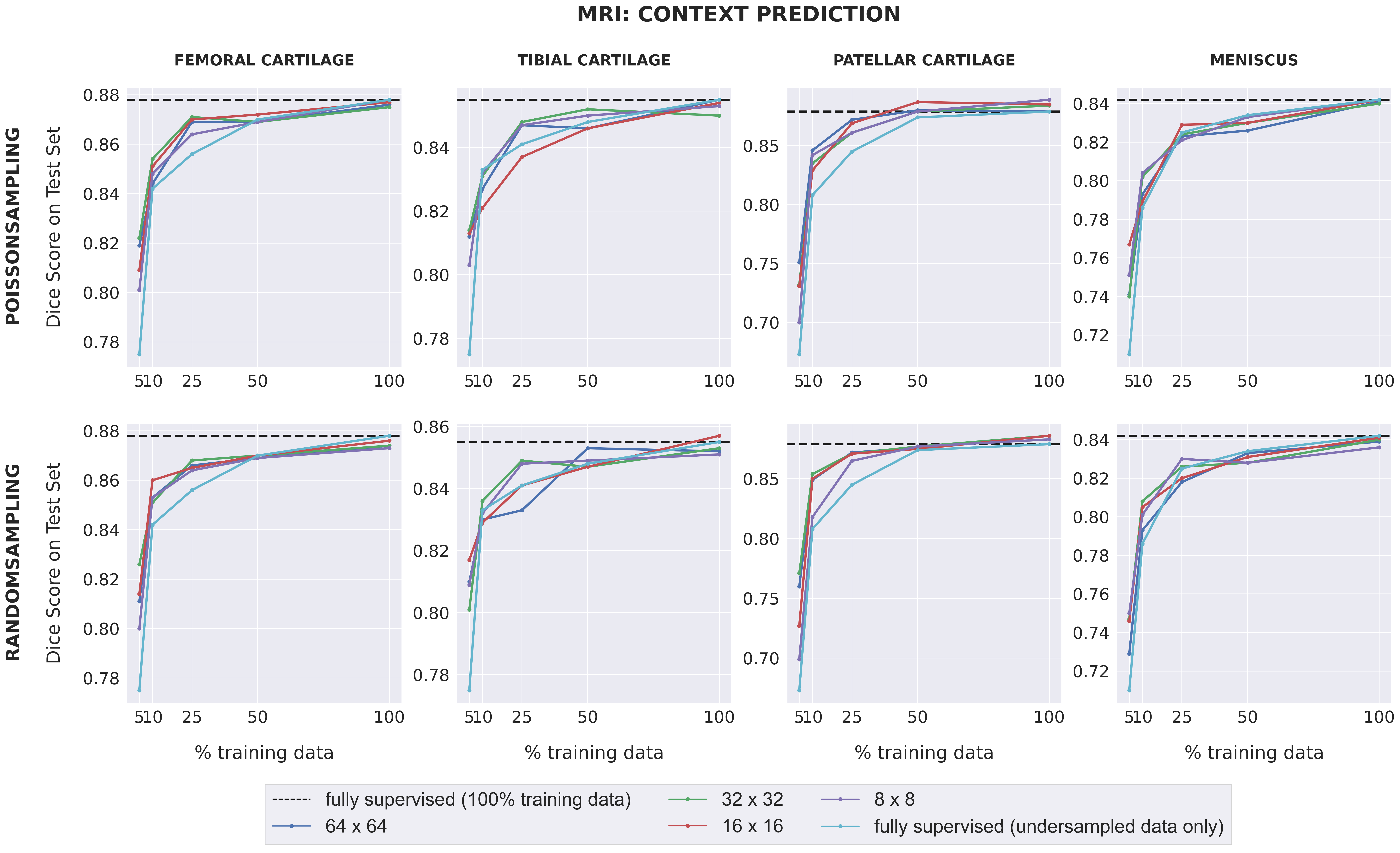}
            \caption{The downstream segmentation performance on the MRI dataset for the Context Prediction pretext task as measured by 
            the Dice score 
            for every combination of 
            patch size and sampling method used during pretraining, evaluated 
            in five different scenarios of training data availability. 
            In each scenario, every model is
            trained for segmentation using one of the 
            five different subsets 
            of training data as described in Section \ref{sec:mri_dataset}. 
            The black dotted line
            in each plot indicates the performance 
            of a fully supervised model trained using all available training images. 
            The light blue 
            curve indicates the performance of a fully supervised model when trained using 
            each of the five 
            different subsets of training data.}
            \label{fig:mri_patch_size_figure_cp}
        \end{figure}
        
        \begin{figure}[ht]
            \centering 
            \includegraphics[width=\textwidth]{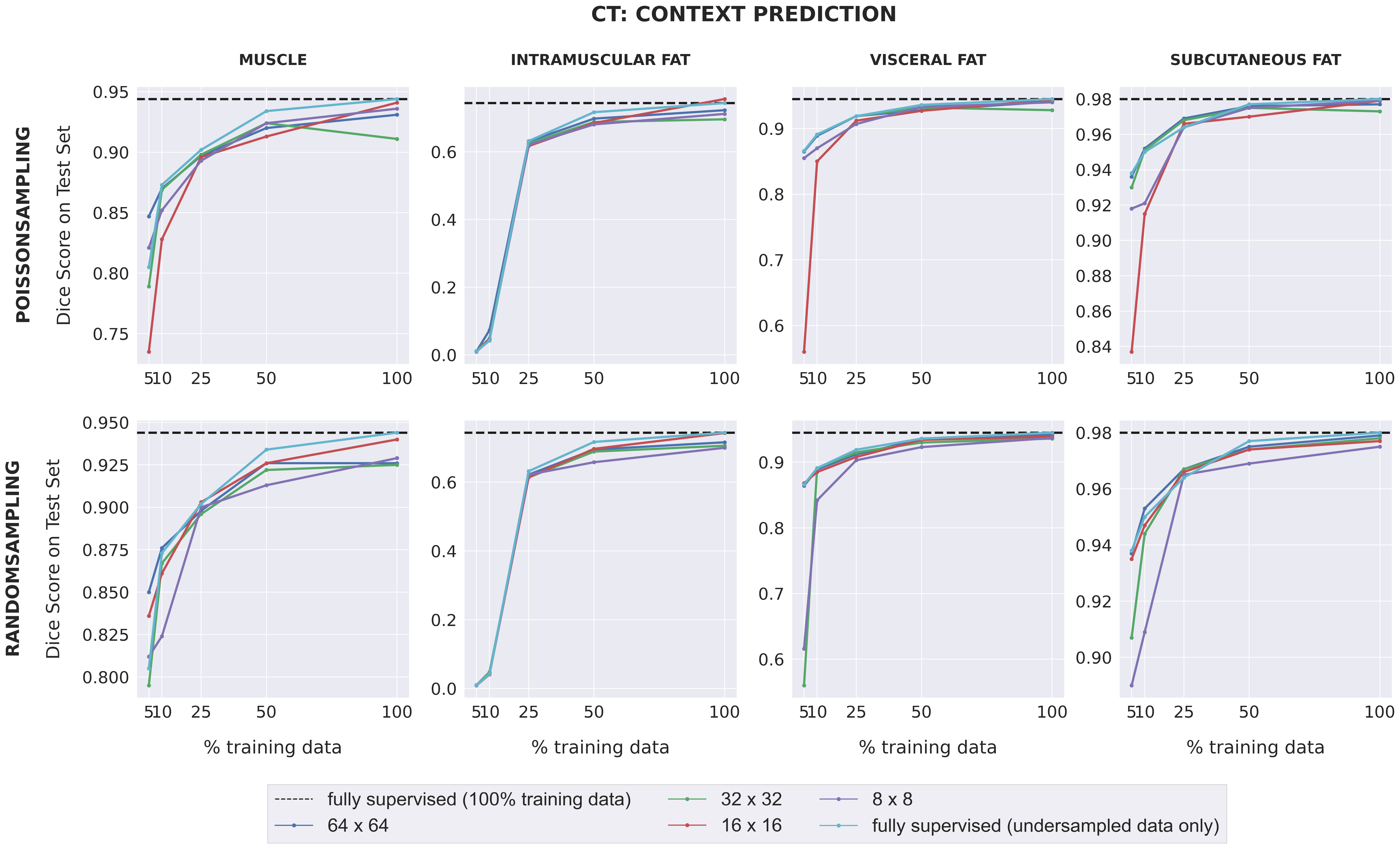}
            \caption{The downstream segmentation performance on the CT dataset for the Context Prediction pretext task as measured by 
            the Dice score 
            for every combination of 
            patch size and sampling method used during pretraining, 
            evaluated in five different scenarios of training data availability. 
            In each scenario,
            every model is trained for segmentation using one of the 
            five different subsets 
            of training data as described in Section \ref{sec:ct_dataset}. 
            The black dotted line 
            in each plot indicates the performance 
            of a fully supervised model trained using all available training images. 
            The light blue 
            curve indicates the performance of a fully supervised model when trained using 
            each of the five 
            different subsets of training data.}
            \label{fig:ct_patch_size_figure_cp}
        \end{figure}
         
\end{document}